# Observation of Anomalous Hall Effect in Bulk Single Crystals of *n*-type Cr-doped Sb₂Te₃ Magnetic Topological Insulator

Ali Sarikhani[a,†,*], Mathew Pollard[b,†], Jacob Cook[c], Sheng Qiu[b], Seng Huat Lee[b], Laleh Avazpour[b], Jack Crewse[b], William Fahrenholtz[d], Guang Bian[c], and Yew San Hor[a,b,*]

[a] Material Research Center, Missouri University of Science and Technology, Rolla, MO 65409

[b] Department of Physics, Missouri University of Science and Technology, Rolla, MO 65409

[c] Department of Physics and Astronomy, University of Missouri, Columbia, MO 65211

[d] Department of Materials Science and Engineering, Missouri University of Science and Technology, Rolla, MO 65409

† Equal Contribution.

* Corresponding Authors: yhor@mst.edu (Y. S. Hor), as5kw@mst.edu (A. Sarikhani)

**The exploration of topological Dirac surface states is significant in the realms of condensed matter physics and future technological innovations. Among the materials garnering attention is Sb₂Te₃, a compound that theoretically exhibits topological insulating properties. However, its inherent *p*-type nature prevents the direct experimental verification of its Dirac surface state due to the Fermi level's alignment with the valence band. In this study, by doping Cr atoms into Sb₂Te₃, *n*-type behavior is observed in the Hall resistance measurements. Remarkably, the Cr-doped Sb₂Te₃ not only shows ferromagnetism with a high transition temperature of approximately 170 K but also exhibits an anomalous Hall effect (AHE). The Cr doping also allows for a controlled method for Fermi level tuning into the band gap. These properties spotlight its potential as an *n*-type magnetic topological insulator (MTI) as well as a material candidate for the quantum anomalous Hall effect (QAHE), opening new avenues for applications in spintronics and quantum devices.**

Three-dimensional topological insulators (TIs) have garnered significant attention over the past decade due to their unique electronic structure, i.e. an insulating bulk coexisting with gapless surface states protected by time-reversal symmetry (TRS) [1,2]. These surface states originate from strong spin-orbit coupling, as depicted in materials like $Sb_2Te_3$, $Bi_2Se_3$ and $Bi_2Te_3$, and are characterized by linearly dispersing Dirac cones that connect the valence and conduction bands, while the bulk remains gapped [3]. Doping TIs with magnetic cations introduces magnetic exchange interactions that can induce a long-range magnetic order, transforming a non-magnetic TI into a magnetic topological insulator (MTI) [4,5]. This magnetic doping not only modifies the electronic structure by shifting the Fermi level, enabling tuning between the valence and conduction bands, but also fundamentally alters the topological protection of the surface states. Specifically, the introduction of magnetism breaks TRS and opens a gap at the Dirac point, thereby destroying the intrinsic spin-momentum locking characteristic of pristine TIs [6].

Direct observation of surface states in topological insulators (TIs) is commonly achieved using angle-resolved photoemission spectroscopy (ARPES), which provides direct evidence of the Dirac cone, particularly in $n$-type systems where the Fermi level lies near or within the bulk band gap [7]. However, in $p$-type materials such as pristine $Sb_2Te_3$, the Fermi level resides deep within the valence band, making it difficult to resolve the surface Dirac states, as they become obscured by the bulk electronic structure. Previous efforts to achieve $n$-type conduction in TIs, such as doping $Bi_2Se_3$ with Mn, Cu, or Ca [8-10], have demonstrated tunability of the charge carrier type, but successfully converting $Sb_2Te_3$ to $n$-type conduction has remained a significant challenge [11-13]. Nevertheless, precise Fermi level tuning through chemical doping is crucial, not only for revealing gapped surface states, but also for enabling quantum phenomena such as the quantum anomalous Hall effect [14-18], which has been recently observed in intrinsic magnetic topological insulators like the septuple-layer $MnBi_2Te_4$ [19].

In this study, we report the first successful realization of $n$-type $Sb_2Te_3$ through Cr doping, as confirmed by Hall resistance measurements. This enables direct experimental verification of the material's topological surface states using ARPES. With the Fermi level shifted into the bulk band gap, the long-concealed Dirac states become accessible, solidifying the identification of $Sb_2Te_3$ as a topological insulator. Moreover, Cr doping introduces robust ferromagnetic ordering with a Curie temperature of approximately 170 K, indicative of strong magnetic exchange interactions between the Cr ions embedded in the host lattice. This magnetic behavior, when combined with the system's intrinsic spin-orbit coupling, gives rise to a pronounced anomalous Hall effect (AHE), a hallmark of magnetic topological systems, in which the spontaneous magnetization acts as an internal pseudo-magnetic field that deflects charge carriers [20,21]. Our ARPES measurements further demonstrate that the Fermi level can be tuned into the gap opened by magnetic exchange interactions at the Dirac point, offering a realistic pathway toward QAHE observation in this system. Taken together, our findings establish Cr-doped $n$-type $Sb_2Te_3$ single crystal as a rare example of a bulk magnetic topological insulator exhibiting both ferromagnetism and AHE. The successful $n$-type tuning, combined with the emergence of robust spin-charge interaction, positions this material as a highly promising candidate for dissipationless spintronic devices and quantum electronic applications [22-26].

The X-ray diffraction (XRD) patterns of $Cr_{0.2}Sb_2Te_3$ and $Cr_{0.2}Sb_2Te_{3-\delta}$ single crystals display a series of sharp and intense (0 0 n) reflections, indicative of highly oriented crystalline growth along the c-axis, as depicted in Figure 1a. The presence of only (0 0 n) peaks confirms the formation of a well-ordered layered structure consistent with the rhombohedral crystal symmetry (space group $R\overline{3}m$) typical of the $Sb_2Te_3$ family. A slight difference in peak positions between the Te-deficient ($Cr_{0.2}Sb_2Te_{3-\delta}$) and stoichiometric ($Cr_{0.2}Sb_2Te_3$) samples may reflect subtle lattice parameter changes due to Te vacancies, which can influence the interlayer spacing and potentially tune the electronic and magnetic properties. As indicated in Figure 1b, each quintuple layer consists of 5 atomic planes stacked along the crystallographic $c$-axis in the sequence Te(1)-Sb-Te(2)-Sb-Te(1). These layers are held together by weak van der Waals forces, allowing for easy cleavage along the basal plane. The Cr atoms can be intercalated in between these quintuple layers without disrupting the intralayer covalent bonding network.

Figure 1c show scanning tunneling microscopy (STM) images at multiple scales for the cleaved surface of $Cr_{0.2}Sb_2Te_3$ single crystals. Figure 1c presents a large-area topographic scan (~200 nm scale) that reveals atomically flat terraces separated by step edges, indicating clean cleavage along the van der Waals gap between quintuple layers. These terraces are consistent with the natural c-axis cleavage plane of the rhombohedral $Sb_2Te_3$ structure. Figure 1c also zooms into a smaller area (~80 × 80 nm²), highlighting the uniform topography and distribution of nanoscale contrast features. These bright and dark spots are likely due to point defects such as substitutional Cr atoms, Cr intercalants, or intrinsic vacancies (Te or Sb), reflecting the dopant incorporation and local stoichiometry variation during crystal growth. The inset in the lower-right corner of Figure 1c displays an atomic-resolution STM image, clearly resolving the surface hexagonal lattice with a periodicity consistent with the (0001) plane of $Sb_2Te_3$. The observed lattice symmetry confirms high crystallinity and long-range structural order at the atomic scale. Minor local distortions in lattice contrast further support the presence of atomic scale inhomogeneities associated with Cr doping.

Figure 2a displays the temperature-dependent DC magnetization ($M_{DC}$) of $Cr_xSb_2Te_3$ single crystals with varying Cr concentrations (x = 0.05, 0.1, 0.15, and 0.2), measured under applied field $H$ = 5 kOe. All samples exhibit a clear ferromagnetic transition near ~170 K, with increasing magnetization magnitude correlating with higher Cr content. This transition is attributed to long-range ferromagnetic ordering induced by Cr doping. For the highest doping level (x = 0.2), an additional gradual increase in magnetization is observed below ~20 K, accompanied by a saturation trend as temperature approaches zero. While the origin of this low-temperature feature remains to be fully clarified, its behavior is distinct from simple paramagnetism and may suggest the presence of additional magnetic interactions or phase inhomogeneity at high Cr concentrations. Further studies are required to determine whether this low-temperature upturn reflects a secondary ferromagnetic contribution, local magnetic ordering, or another mechanism.

Figure 2b presents isothermal magnetization curves ($M_{DC}$ vs. $\mu_0 H$) for $Cr_{0.2}Sb_2Te_3$ single crystals measured at 2 K and 5 K under different field orientations. At both temperatures, clear hysteresis loops are observed when the magnetic field is applied along the $c$-axis (red and blue curves), indicating ferromagnetic ordering with a coercive field and remanent magnetization. In contrast, when the field is applied in the $ab$ plane at 5 K (green curve), the magnetization exhibits a nearly linear, non-hysteretic response, suggesting that the $ab$ direction behaves as a magnetic hard axis. At 2 K, when the magnetic field is applied along the $c$-axis (blue curve), the hysteresis loop becomes more pronounced compared to the 5 K measurement (red curve), consistent with stronger ferromagnetic order at lower temperatures. This anisotropic behavior confirms the presence of uniaxial magnetic anisotropy in $Cr_{0.2}Sb_2Te_3$, with the magnetic easy axis oriented along the c-axis, perpendicular to the layered crystal planes. Such behavior is characteristic of Cr-doped topological insulators and has been similarly observed in Cr-doped $Bi_2Se_3$ systems [27]. It arises from the interplay between strong spin-orbit coupling and the intrinsic structural anisotropy of the van der Waals-layered lattice, which favors out-of-plane spin alignment. The pronounced difference in magnetization response between the c-axis and in-plane directions further supports this magnetic anisotropy, confirming that the easy axis lies perpendicular to the basal plane [28].

The metallic behavior observed in Cr-doped $Sb_2Te_3$ is consistent with the ferromagnetic ordering evidenced in the magnetization results. To further explore the transport characteristics, Figure 3a shows the temperature dependence of the longitudinal resistivity ($\rho_{xx}$) for $Cr_{0.2}Sb_2Te_3$ and Te-deficient $Cr_{0.2}Sb_2Te_{3-\delta}$ single crystals. Both samples exhibit a gradual decrease in resistivity with decreasing temperature, indicating metallic conduction enabled by Cr-induced charge carriers, suggesting that the Fermi level is tuned into or near the conduction band. Notably, below approximately 20 K, the resistivity shows a subtle deviation from the monotonic metallic trend with a slight flattening or reduced slope. This feature may reflect enhanced scattering effects at low temperatures, possibly arising from magnetic impurities (Cr ions), structural disorder, or weak localization phenomena. In magnetically doped topological insulators, such low-temperature deviations are frequently attributed to interactions between itinerant carriers and localized magnetic moments, or to increased quantum interference effects. These behaviors highlight the complex interplay between magnetic and electronic degrees of freedom in Cr-doped $Sb_2Te_3$.

Following the metallic behavior observed in the longitudinal resistivity (Figure 3a), we further examined the carrier type and magnetic response through Hall resistivity ($R_{xy}$) measurements. Figures 3b and 3c show the transverse Hall resistivity as a function of magnetic field for $Cr_{0.2}Sb_2Te_3$ and $Cr_{0.2}Sb_2Te_{3-\delta}$, respectively, across a range of temperatures from 2 to 150 K. The insets present full-field measurements up to $\pm 5$ T, while the main panels highlight the low-field regime where the anomalous Hall effect (AHE) emerges. At low temperatures, both samples exhibit a clear hysteresis in $R_{xy}$ that saturates around $\pm 0.05$ T, consistent with the coercive field seen in the magnetization data (Figure 2b). This confirms the presence of ferromagnetism and a strong AHE contribution, which follows the relation $R_{xy} = R_0 B + R_s M$. Here, the anomalous term ($R_s M$) dominates at low fields due to magnetization, while the ordinary Hall term ($R_0 B$) becomes evident at higher fields [29,30]. In this equation, $R_0$ is ordinary Hall coefficient while $R_s$ represents the anomalous Hall coefficient. The close agreement between the coercive field in the Hall and magnetization data further supports this interpretation. Most notably, the sign of the high-field slope of $R_{xy}$ reveals the dominant carrier type. In stoichiometric $Cr_{0.2}Sb_2Te_3$ (Figure 3b), the slope is positive, indicating $p$-type (hole-like) conduction. In contrast, the Te-deficient $Cr_{0.2}Sb_2Te_{3-\delta}$ sample (Figure 3c) exhibits a negative slope, demonstrating a transition to $n$-type (electron-like) carriers. This change is attributed to the introduction of Te vacancies, which act as donor defects, providing extra electrons and shifting the Fermi level toward or into the conduction band.

This Fermi level tuning is particularly significant, as $Sb_2Te_3$ is inherently $p$-type, and achieving stable $n$-type conductivity has been a longstanding challenge. To our knowledge, this represents one of the first demonstrations of $n$-type behavior in bulk Cr-doped $Sb_2Te_3$, made possible through controlled Te deficiency. Such Fermi level tuning is essential for aligning it near the Dirac point and enabling magnetic gap opening in the topological surface states below the Curie temperature, which is a prerequisite for realizing the quantum anomalous Hall effect. Unlike other systems such as $Bi_2Se_3$, where gate voltage provides a direct handle on Fermi level tuning, the carrier density in Cr-doped $Sb_2Te_3$ appears to be governed by a complex interplay among Cr incorporation, Sb/Te site occupancy, and resulting defect chemistry. While the precise relationship remains to be determined, our data suggest that compositions with increased Te vacancies are most effective in pushing the system toward $n$-type behavior.

The variability in carrier type and Fermi level position observed in our Hall measurements is further examined via angle-resolved photoemission spectroscopy (ARPES). Figure 4a shows the ARPES spectrum of a representative $Cr_{0.2}Sb_2Te_3$ sample, overlaid with DFT-calculated bulk band structure. The spectrum displays characteristic energy-momentum dispersion of the magnetic topological insulator, with a partially visible topological surface state (TSS) near the $\Gamma$-point. The Fermi level lies below the Dirac point and within the valence band, consistent with the p-type character revealed by Hall measurements (Figure 3b). By systematically varying Sb and Cr concentrations, we achieve a Fermi level tuning range of approximately 0.057 eV. For example, increasing Sb content raises the Fermi level by ~0.038 eV relative to the $Cr_{0.2}Sb_2Te_3$ baseline, moving it closer to the Dirac point. Although the Dirac point is not yet fully reached, these adjustments bring the Fermi level within or near the bulk band gap, suggesting that further optimization could enable direct observation of a magnetically gapped surface state for realizing the quantum anomalous Hall effect.

Figures 4b illustrates the electronic structure of the higher Fermi level sample, i.e., the Te-deficient $Cr_{0.2}Sb_2Te_{3-\delta}$, which exhibits n-type behavior in Hall measurements (Figure 3c). Surprisingly, the ARPES spectrum of $Cr_{0.2}Sb_2Te_{3-\delta}$ appears nearly identical to that of the stoichiometric sample, showing no clear Fermi level shift into the conduction band. This discrepancy between ARPES and transport measurements likely stems from the surface-sensitive nature of ARPES, which probes only the top few atomic layers. In Te-deficient samples, surface band bending or depletion effects may pin the Fermi level closer to the valence band, even when the bulk exhibits n-type conduction. Additionally, time-dependent surface degradation or sample inhomogeneity such as Cr segregation seen in the STM image (Figure 1d) may further obscure subtle changes in the electronic structure. Despite this inconsistency, the overall trend of Fermi level tuning remains evident, and Hall measurements reliably indicate bulk carrier type. Addressing surface-bulk discrepancies through improved surface preparation, gated ARPES, or bulk-sensitive spectroscopies could provide more definitive confirmation in future studies. It is also worth noting that the ARPES measurements were conducted at liquid nitrogen temperature (109 K). In contrast, the Hall resistance data for both the Cr-intercalated and Te-deficient samples exhibit linear behavior at around this temperature, without pronounced p- or n-type hysteresis features. Future studies could explore performing ARPES at lower temperatures to better capture the electronic environment associated with n-type conduction.

Furthermore, the AHE observed in Figures 3b and 3c is consistent with ferromagnetic hysteresis seen in magnetization (Figure 2b) and the electronic structure inferred from ARPES. At 2 K, the coercive field (~0.02 T) marks the field at which net magnetization switches direction, canceling the transverse Hall voltage due to equal contributions from spin-up and spin-down carriers. As the field increases past the saturation point (~0.05 T), full spin polarization is achieved, producing a maximum in $R_{xy}$. This symmetry in AHE hysteresis under positive and negative magnetic fields confirms the magnetic origin of the Hall signal and its proportionality to spontaneous magnetization. Together, these results provide a coherent picture of magnetic and electronic behavior in Cr-doped $Sb_2Te_3$ and demonstrate the material's potential for tunable topological transport [31].

In this study, we demonstrate the observation of an intrinsic AHE in bulk single crystals of Cr-doped $Sb_2Te_3$, confirming that ferromagnetic ordering and topological transport features can coexist in the bulk of this magnetic topological insulator system. By reducing the Te stoichiometry to $Cr_{0.2}Sb_2Te_{3-\delta}$, we successfully converted the material from $p$-type to $n$-type conduction, as confirmed by Hall measurements. While ARPES did not reveal a clear Fermi level shift into the conduction band for the Te-deficient sample, this discrepancy is likely due to the surface sensitivity of ARPES, in contrast to the bulk nature of transport measurements. Cr doping introduced ferromagnetism with a Curie temperature around 170 K, which supports the emergence of the AHE. Although the Dirac point was not directly accessed, we show that the Fermi level can be tuned closer to the bulk band gap by adjusting the elemental composition, particularly Te deficiency. These results establish that AHE can be robustly realized in bulk Cr-doped $Sb_2Te_3$ single crystals and underscore the potential of this system for exploring magnetic topological transport.

## Methods

### Synthesis

High-quality single crystals of Cr-doped $Sb_2Te_3$ were synthesized using a modified Bridgman growth method via slow cooling in evacuated quartz ampoules. Stoichiometric amounts of high-purity elemental chromium (Cr, 99.999%), antimony (Sb, 99.999%), and tellurium (Te, 99.999%) were carefully weighed in a glovebox to minimize surface oxidation. The typical target compositions used were $Cr_xSb_2Te_3$ with x = 0.05-0.2, including both stoichiometric and Te-deficient variants, i.e., $Cr_{0.2}Sb_2Te_{3-\delta}$. The elemental mixtures were loaded into clean, dry quartz ampoules with 10 mm inner diameter and approximately 10 cm in length. The ampoules were thoroughly evacuated ($\sim 10^{-6}$ mbar) and flame-sealed under high vacuum to prevent contamination and oxidation during the high-temperature process. To further ensure cleanliness and promote uniform mixing, the ampoules were flushed multiple times with ultra-high-purity argon gas before final evacuation.

The sealed ampoules were placed in a box furnace and heated gradually to 950 °C at a rate of 0.5 °C/min. The mixtures were maintained at this temperature for 24 hours to allow for complete and homogeneous mixing of the constituent elements. Periodic manual rocking or spinning of the ampoules was performed during the period to promote uniform distribution and reaction of Cr with the Sb-Te matrix. After homogenization, the precursor was cooled slowly to room temperature at a rate of 0.5 °C/min to promote the formation of large, well-ordered single crystals. Slow cooling is crucial for minimizing crystal defects and phase separation, particularly for doped systems with narrow solidification windows. Upon reaching room temperature, the ampoules were opened, and the resulting crystals were mechanically cleaved to expose shiny lamellar flakes along the basal (0001) plane, typical of the rhombohedral $Sb_2Te_3$ structure. The as-grown crystals were gray, shiny, and easily cleavable, with thicknesses ranging from hundreds of microns to several millimeters. All samples (with nominal compositions of $Cr_xSb_2Te_3$ and $Cr_{0.2}Sb_2Te_{3-\delta}$ with $\delta=0.1$) were stored in an inert atmosphere to minimize surface oxidation prior to characterization.

### X-ray Diffraction

The crystal structures of $Cr_{0.2}Sb_2Te_{3-\delta}$ and $Cr_{0.2}Sb_2Te_3$ were characterized using a PANalytical X'Pert Multi-Purpose Diffractometer equipped with a copper source ($\lambda = 0.15418$ nm) and a minimum step size of $2\theta = 0.06°$. All major peaks in the XRD patterns of $Cr_{0.2}Sb_2Te_{3-\delta}$ and $Cr_{0.2}Sb_2Te_3$ correspond to the cubic lattice of antimony telluride ($Sb_2Te_3$). The peaks indexed as (0 0 n), where n = 3, 6, 9…, are clearly indicated in Figure 1a. As shown in the XRD patterns, no impurity phases are detected in either $Cr_{0.2}Sb_2Te_{3-\delta}$ or $Cr_{0.2}Sb_2Te_3$. The c-axis lattice parameter of undoped $Sb_2Te_3$ decreases from 30.458 Å to 30.419 Å for $Cr_{0.2}Sb_2Te_{3-\delta}$, and to 30.402 Å for $Cr_{0.2}Sb_2Te_3$. These values indicate only slight changes compared to the undoped $Sb_2Te_3$ structure. Lattice parameters were determined using the Rietveld refinement method, which is widely employed for analyzing ion-substituted crystal systems.

### Transport Measurements

A Quantum Design Physical Property Measurement System (PPMS) was used throughout this study. Magnetization as a function of temperature and magnetic field was measured using the DC magnetization module of the PPMS. Hall resistance measurements were conducted using a five-probe configuration under the ACT option, and the Hall resistance for the bulk material was corrected by factoring in the sample thickness (t), which corresponds to the lamellar crystal thickness along the c-axis.

### ARPES-STM Measurements

The Cr-doped $Sb_2Te_3$ single crystals were prepared for STM and ARPES measurements by in-situ cleaving the surface in an integrated MBE-STM ultrahigh vacuum (UHV) system with base pressure below $2 \times 10^{-10}$ mbar. Surface topography mappings were performed in situ by using the SPECS Aarhus-150 scanning tunneling microscopy (STM) system with a Tungsten tip at room temperature. Angle Resolved Photo-emission Spectroscopy (ARPES) measurements were performed in situ at liquid nitrogen temperatures (109 K) using a SPECS PHOIBOS-150 hemisphere analyzer with a SPECS UVS 300 helium discharge lamp (He I$\alpha$ = 21.2 eV). The energy resolution is 40 meV at 109 K. Characteristic, high intensity electron bands at $\Gamma$ were quantified using a Gaussian Multi-peak fitting (GMPF) of the summed intensity profile $\pm$ 0.05 Å$^{-1}$ around the zone center, as implemented in the Igor Pro Multi-peak fitting software package. Smoothing factors were increased in the GMPF until the characteristic intensity peak at $\sim$ -0.2 eV consolidated into a singular peak feature.

### DFT Calculations

All theoretical band structure and first-principles calculations were performed with density functional theory (DFT) as implemented in the Vienna ab initio Simulation Package (VASP). We used the Perdew-Burke-Ernzerhof (PBE) form for the exchange-correlation functional with a plane-wave cut-off energy of 300 eV. Spin-Orbit Coupling (SOC) was included in all calculations.

## Competing interests

The authors declare no competing interests.

## Acknowledgments

The authors thank Dr. Eric Bohannan for helpful support and discussion on XRD measurements. The work was partially supported by a grant from NSF DMR-1255607. Materials Research Center at Missouri S&T sponsored in part for pursuing this research.

**Figure Captions**

**Figure 1**. **a** X-ray diffraction (XRD) patterns of $Cr_{0.2}Sb_2Te_{3-\delta}$ and $Cr_{0.2}Sb_2Te_3$ single crystals, showing the (0 0 n) reflections of the rhombohedral $Sb_2Te_3$ structure.

**b** Schematic illustration of the rhombohedral unit cell with Cr intercalation between the quintuple van der Waals layers.

**c** Scanning tunneling microscopy (STM) image of the cleaved surface, highlighting atomic-level topography.

**Figure 2**. **a** Temperature-dependent magnetization (M-T) for samples with x = 0.05, 0.1, 0.15, and 0.2, revealing a ferromagnetic transition near 170 K. Lower x samples show suppressed magnetic response but similar transition temperature. **b** Isothermal DC magnetization of Cr-doped $Sb_2Te_3$ measured at 2 K and 5 K with magnetic field applied parallel and perpendicular to the ab plane, showing clear ferromagnetic hysteresis out-of-plane.

**Figure 3**. **a** Longitudinal resistivity ($\rho_{xx}$) as a function of temperature from 5 K to 300 K, illustrating metallic behavior with Cr doping. **b** Hall resistance ($R_{xy}$) vs magnetic field for $Cr_{0.2}Sb_2Te_3$ at various temperatures, displaying p-type anomalous Hall effect (AHE) hysteresis below the Curie temperature. **c** Hall resistance for Te-deficient $Cr_{0.2}Sb_2Te_{3-\delta}$, indicating a transition to n-type behavior.

**Figure 4**. ARPES spectra for the two representative samples at liquid nitrogen temperatures (109 K): **a** $Cr_{0.2}Sb_2Te_3$, **b** $Cr_{0.2}Sb_2Te_{3-\delta}$. Band dispersions along the Γ-K and Γ-M directions show systematic shifts in Fermi level with doping. Fermi level positions are tuned within a range of ~0.057 eV, approaching the Dirac point in panel.

# Figure 1

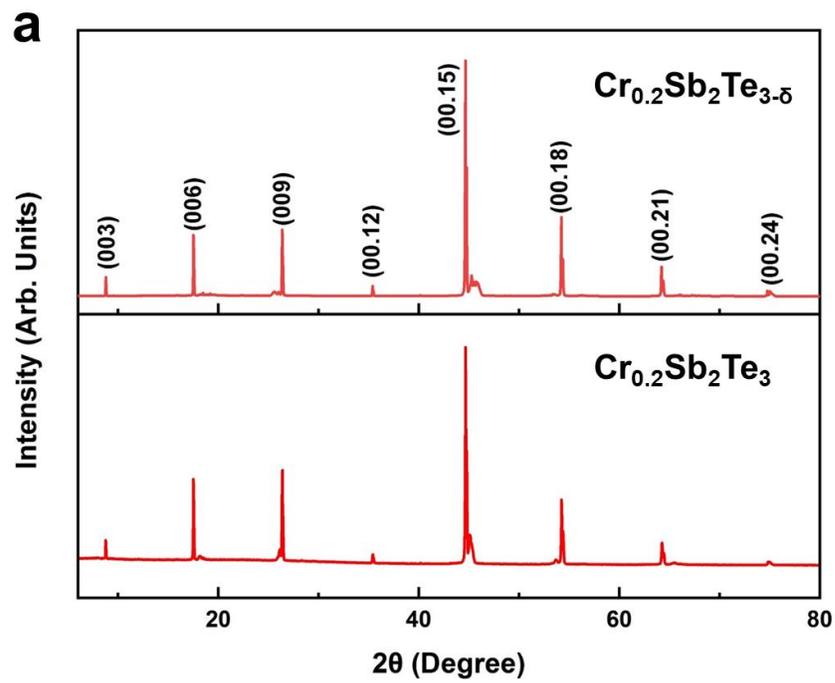

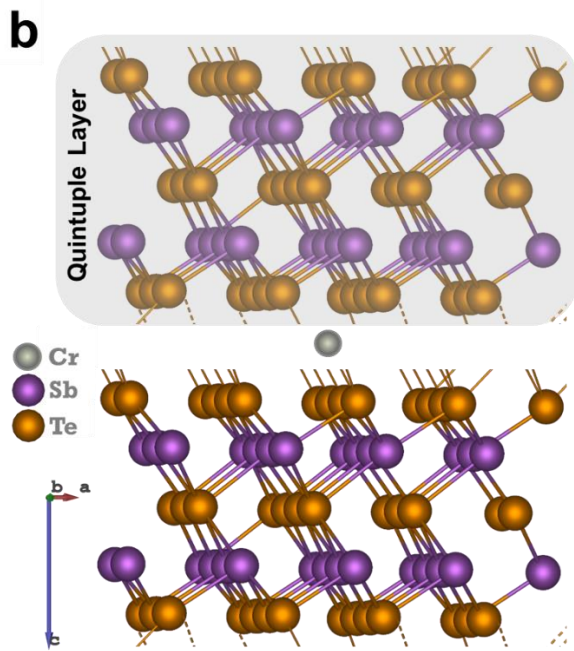

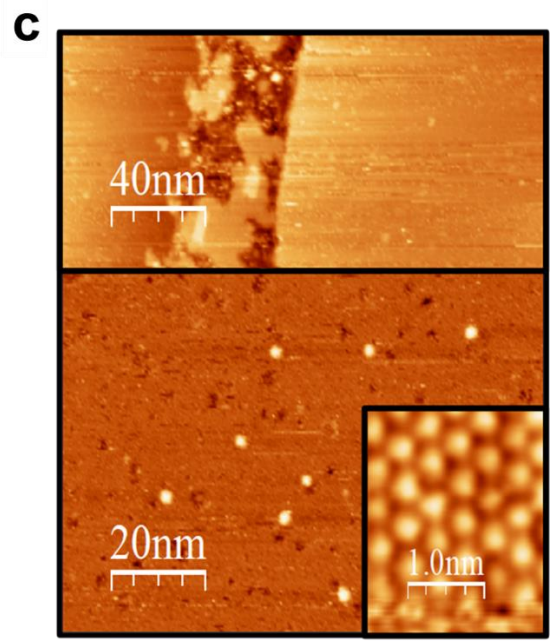

**Figure 2**

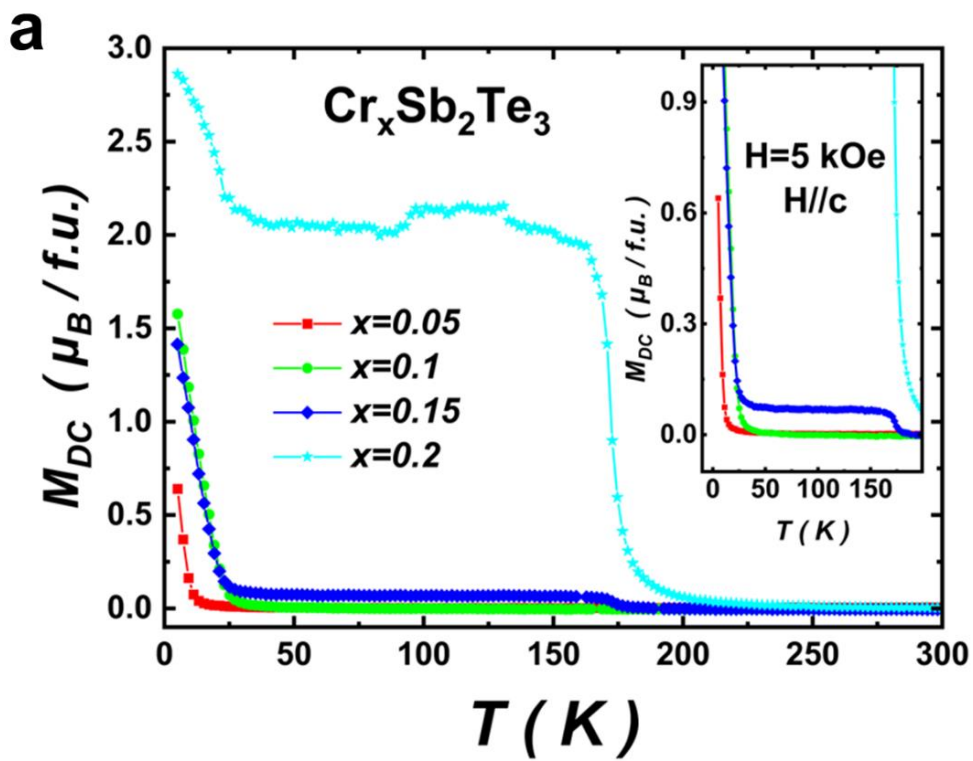

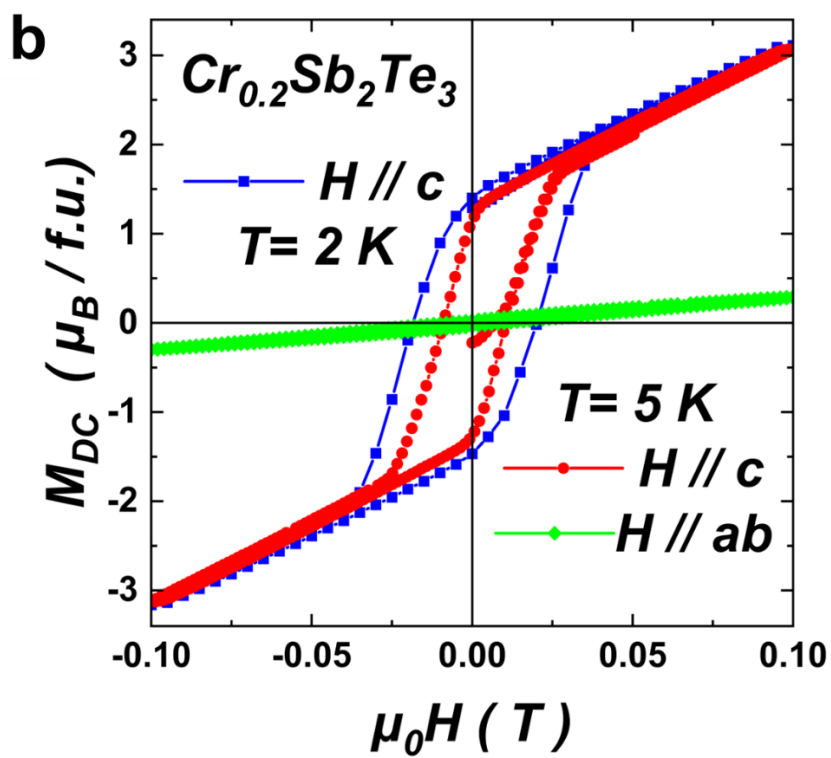

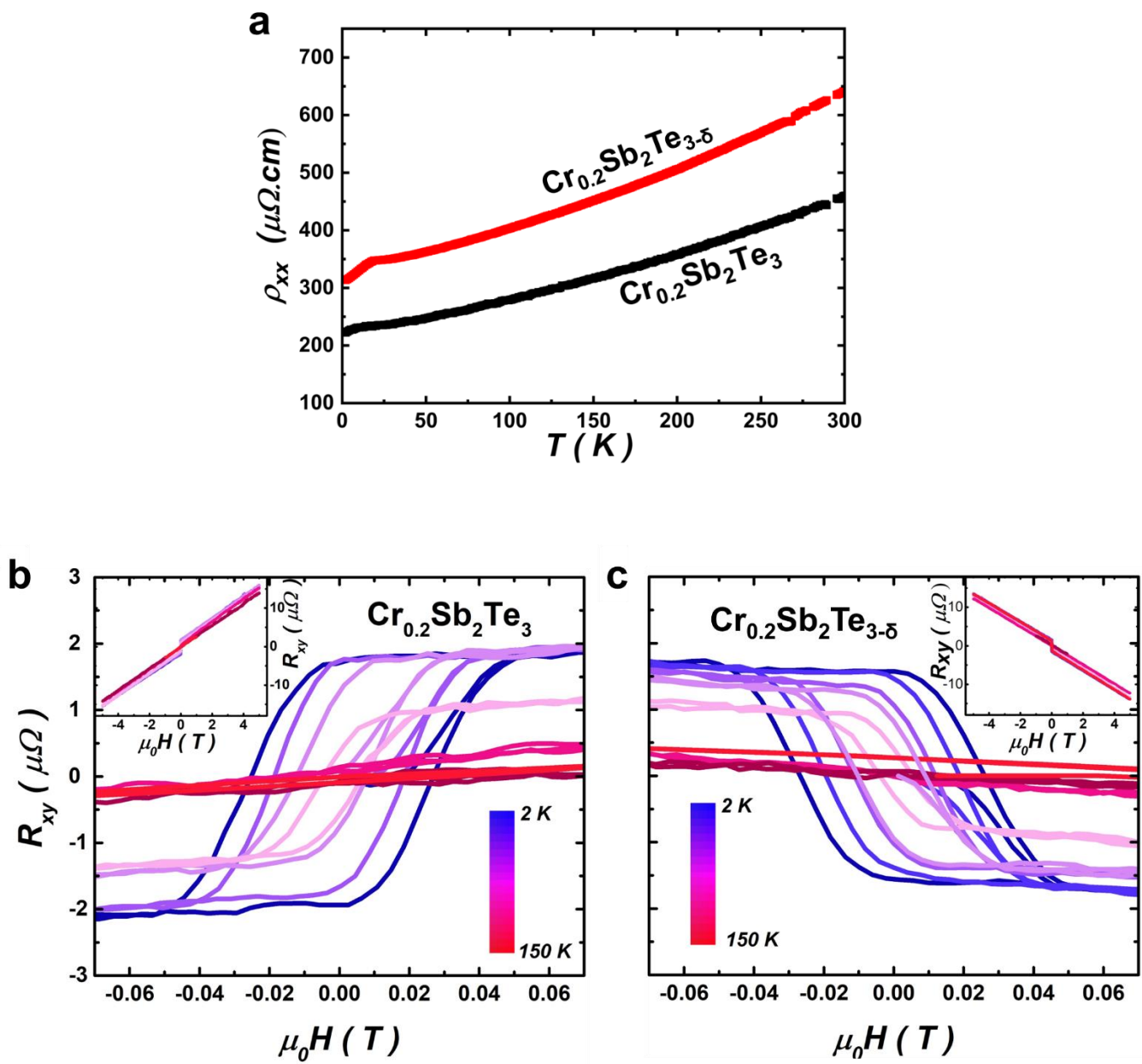

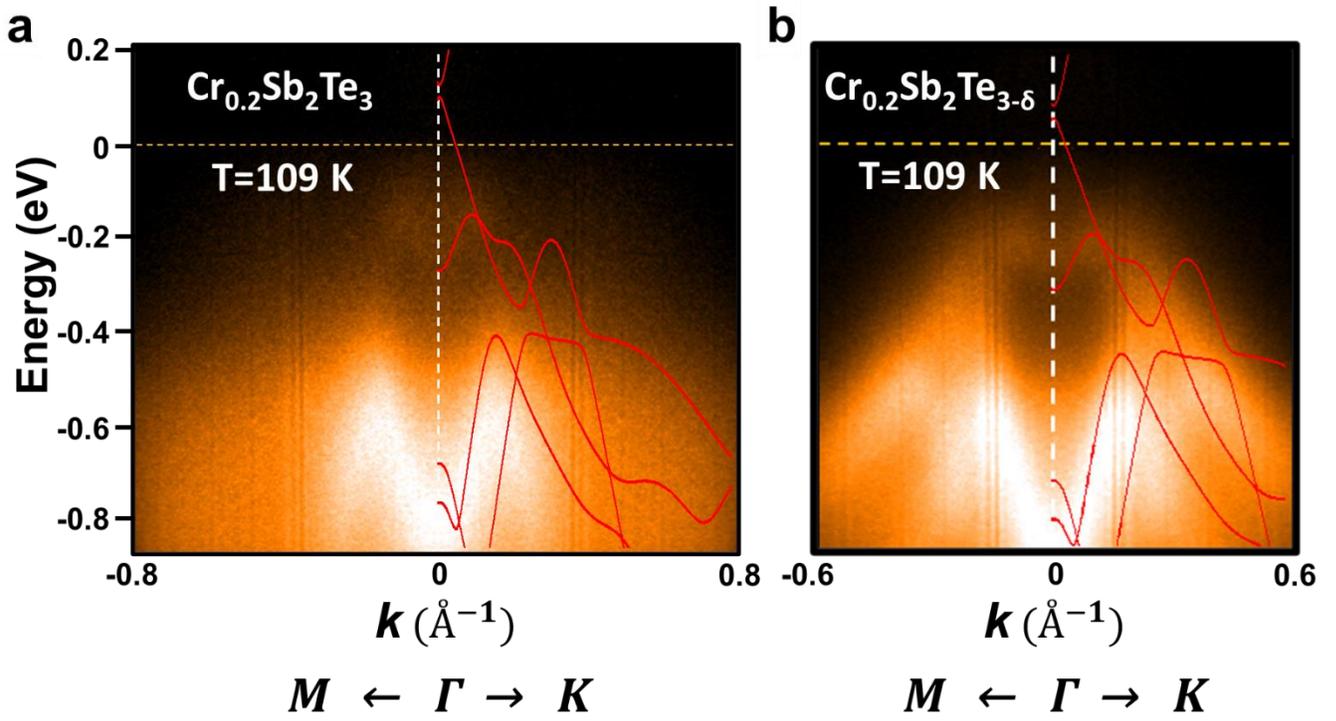